\documentclass{article}
\pdfoutput=1
\usepackage{amsfonts,amsmath,color}
\usepackage{graphicx}
\usepackage{graphics}
\usepackage{subcaption}
\usepackage{epsfig}
\usepackage{amssymb}
\usepackage{jheppub}
\begin{document}
\title{Fermionic Halos at Finite Temperature in AdS/CFT}
\author[a,c]{Carlos R. Arg\"uelles}
\author[b,c]{Nicol\'{a}s E. Grandi}
\affiliation[a]{Facultad de Ciencias Astron\'omicas y Geof\'\i sicas de La Plata \\ Observatorio Astron\'omico, Paseo del Bosque s/n, B1900FWA La Plata, Argentina}
\affiliation[b]{Instituto de F\'{\i}sica de La Plata - CONICET,\\ Calle 49 y 115 s/n, C.C. 67, 1900 La Plata, Argentina}
\affiliation[c]{Departamento de F\'{\i}sica - UNLP,\\ Calle 49 y 115 s/n, C.C. 67, 1900 La Plata, Argentina}
\emailAdd{Carlos.Arguelles@icranet.org}
\emailAdd{grandi@fisica.unlp.edu.ar}

\abstract{
We explore the gravitational backreaction of a system consisting in a very large number of elementary fermions at finite temperature, %\textbf{and for different degrees of central degeneracy},
in asymptotically AdS space. We work in the hydrodynamic approximation, and solve the Tolman-Oppenheimer-Volkoff equations with a perfect fluid whose equation of state takes into account both the relativistic effects of the fermionic constituents, as well as its finite temperature effects. We find a novel {\it dense core-diluted halo} structure for the density profiles in the AdS bulk, similarly as recently reported in flat space, for the case of astrophysical dark matter halos in galaxies. We further study the critical equilibrium configurations above which the core undergoes gravitational collapse towards a massive black hole, and calculate the corresponding critical central temperatures, for two qualitatively different central regimes of the fermions: the diluted-Fermi case, and the degenerate case.
As a probe for the  dual CFT, we construct the holographic two-point correlator of a scalar operator with large conformal dimension in the worldline limit, and briefly discuss on the boundary CFT effects at the critical points.
}
\maketitle
\section{Introduction}
\label{sec:introduction}
The subject of fermions at strong coupling is an open field of theoretical research in present day physics. Its importance spans from the description of the strange metal phase in high $T_c$ superconductors, to the behavior of the quark-gluon plasma in high-energy experiments.

In this context, holography has revealed itself as a very useful tool to understand universal properties of strongly coupled fermionic systems. Remarkably, the theoretical results of this approach share many features with the phenomenological knowledge, even if the planar limit required for holography to work (that entails a large number of degrees of freedom at any spacetime point) is not realized in phenomenological setups.

%\textbf{[To me this paragraph should be further detailed to get a better idea of what is happening between Bulk and Boundary.]}
In the holographic description, fermionic operators in the boundary field theory are described by dual fermionic fields in the bulk. The ground state
%\textbf{[of the boundary fields?]} del sistema, es uno solo
is represented by a gravitational background which, according to the limit under study, may or may not take
%\textbf{gravitational}
backreaction effects into account.
%(\textbf{see e.g. \cite{Arsiwalla:2011} for the case of fully degenerate fermions}).
Fermionic excitations correspond to Dirac spinors propagating in such background, while bosonic ones are described by scalar or vector fields.

In the literature, charged fermionic excitations on non-backreactig backgrounds were explored in
%\textbf{[Only two refs.?, if there are more let's put e.g. to these]}
\cite{Faulkner:2009wj}-\cite{Lee:2008xf}. The backreacting case was first investigated in \cite{deBoer:2009wk}-\cite{Arsiwalla:2011} for neutral fermions at zero temperature in global AdS. The backreaction is considered by means of the energy momentum tensor of a perfect fluid, representing the fermionic background. The resulting Tolman-Oppenheimer-Volkoff equations are solved in an asymptotically AdS setup. The case of charged particles at zero temperature in the Poincare patch, more relevant to condensed matter physics, was first introduced in \cite{Hartnoll:2010gu} by similar methods, and further explored in \cite{Hartnoll:2011dm}-\cite{Cubrovic:2009ye}. Its finite temperature extension was investigated in \cite{Hartnoll:2010ik}-\cite{Puletti:2010de}, in an approximation in which Tolman-Oppenheimer-Volkoff equations are solved with a zero-temperature fermionic fluid, and temperature is introduced by means of a black hole horizon.

The issue of Tolman-Oppenheimer-Volkoff equations for neutral fermions in thermodynamic equilibrium at finite temperature,
%\textbf{and fulfilling the Klein thermodynamic equilibrium conditions},
was first investigated in \cite{Gao} in asymptotically flat backgrounds. More recently, its relevance for the physics %and \textbf{astrophysics}
of fermionic dark matter halos in galaxies
%and their central massive black holes
was reported \cite{Bilic2002}-
%\cite{Arguelles2013}-\cite{ArguellesJKPS2014}-\cite{Ruffini2015}-\cite{ArguellesJCAP2016}-
\!\cite{ArguellesMNRAS2016}. It was shown that in a region of parameters a \emph{dense core-diluted halo} structure appears in the density profile. The solution has a high density peak, {\em i.e.} the  ``core'', at its center, suddenly decreasing into a low density plateau, the ``halo'', at the periphery. While the dense core structure is maintained against self-gravity by the degeneracy pressure, the halo holds it own weight by thermal pressure. Such structures allow for a good description of the dark matter halos when contrasted with baryonic data of galaxies, providing at the same time an alternative to the standard picture of a supermassive black hole sitting at the galactic nucleus \cite{Ruffini2015}-\cite{ArguellesMNRAS2016}. Further extensions of such a theoretical model but including fermion self-interactions was introduced in \cite{ArguellesJCAP2016}.
% within an electro-weak like formalism.

It is the aim of this paper to start the investigation of the aforementioned dense core-diluted halo structure in fermionic density profiles in asymptotically AdS backgrounds, and its implications in the holographic context. We focus in the simplest case of a neutral fermionic fluid in $(3+1)$-dimensional global AdS, first investigated in \cite{deBoer:2009wk} for the particular case of zero temperature. We probe the resulting solutions by calculating a scalar correlator of the dual field theory in the world-line approximation.

The paper is organized as follows. In section \ref{sec:bulk} we write the bulk equations for a self-gravitating fluid at finite temperature. In section \ref{sec:correlators} we describe a simple way to obtain correlators of a scalar operator in the worldline approximation. In section \ref{sec:results} we describe the results of our numerical approach. Finally in section \ref{sec:concusions} we summarize our conclusions.

\section{The bulk: self gravitating fermionic fluid at finite temperature}
\label{sec:bulk}
The system under study corresponds to a very large number of neutral self-gravitating fermions in thermodynamic equilibrium, treated within a $3+1$ asymptotically AdS spacetime. The Einstein equations for such system read
\begin{equation}
G_{\mu\nu}+\Lambda g_{\mu\nu}= 8\pi G\, T_{\mu\nu}\,,
\label{eq:einstein-crudas}
\end{equation}
being $\Lambda$ a negative cosmological constant, that can be written in terms of the AdS length $L$ as $\Lambda=-3/L^2$, and $G$ the gravitational constant\footnote{Throughout this paper we work in natural units $\hbar=c=k_B=1$.}. Regarding the fermion energy-momentum tensor, we approximate it as a perfect fluid
\begin{equation}
T_{\mu\nu}=P g_{\mu\nu}+(\rho+P)u_\mu u_\nu\,.
\label{eq:energy-momentum}
\end{equation}
Since we are interested in the limit $mL\gg 1$ in which there is a huge amount of particles within one AdS radius \cite{deBoer:2009wk}, the density and pressure are given by the equations %
\begin{align}
\rho &= \frac{g}{8\pi^3}\int f(p)\sqrt{p^2+m^2 }\,d^3p\,,
\label{eq:rho}\\
P &= \frac{g}{24\pi^3}\int
    f(p)\frac{p^2}{\sqrt{p^2+m^2 }} \,d^3p,
\label{eq:p}
\end{align}
here $g$ is the number of fermionic species (or the spin degeneracy), the integration runs over all momentum space,
and the function $f(p)$
\begin{equation}
f(p)=\frac1{e^{\beta\left(\sqrt{p^2+ m^2}-\mu\right)}+1} \,,
\label{eq:fermi-dirac-distribution}
\end{equation}
is the Fermi-Dirac distribution function for a fermion of mass $m$, where $1/\beta=T$ is the local temperature and $\mu$ is the local chemical potential. This expression sets a double parametric dependence of the density $\rho$ and pressure $P$ on the temperature $T$ and chemical potential $\mu$, which in turn depend on the metric as explained below.

As we will deal with equilibrium configurations of neutral fermions in global AdS, we consider an stationary spherically symmetric metric with the form
\begin{equation}
ds^2=L^2\left(-e^{\nu(r)}\,dt^2+e^{\lambda(r)}\,dr^2+r^2\,d\Omega^2\right)\, ,
\label{eq:metricAdS}
\end{equation}
with $d\Omega^2=d\vartheta^2+\sin^2{\vartheta}d\varphi^2$. In this setup, the local temperature and chemical potential are  radial functions. They are defined by the thermodynamic equilibrium conditions of Tolman  $e^{\nu(r)/2} T=$constant, and Klein  $e^{\nu(r)/2} \mu=$constant, respectively.

\bigskip

In what follows we find convenient to re-scale the temperature and chemical potential as the dimensionless combinations $\tilde \mu=\mu/m$ and $\tilde T=T/m={1}/{\tilde \beta}$. The Tolman condition is then solved by
\begin{equation}
\tilde T=\tilde T_0 e^{\frac{\nu_0-\nu(r)}{2}}\, ,
\label{eq:tolman}
\end{equation}
in terms of the re-scaled temperature $\tilde T_0$ and metric component $\nu_0$ measured at a reference point. The Klein condition can then be rewritten as
\begin{equation}
\tilde \mu=\tilde \mu_0 e^{\frac{\nu_0-\nu(r)}{2}}\, ,
\label{eq:klein}
\end{equation}
where $\tilde \mu_0$ is the value of the re-scaled chemical potential at the same reference point.

Further re-scalings of the
%mass as $\tilde m= Gm/l$, and of the
energy density and pressure as $\tilde \rho=GL^2\rho $ and $\tilde P= GL^2P$ gives the result
\begin{align}
\tilde\rho &= \gamma^2\int_1^\infty \frac{\epsilon^2\sqrt{\epsilon^2-1}}{e^{\tilde \beta\left(\epsilon-\tilde \mu\right)}+1}\,d\epsilon\,,
\label{eq:rhoe}\\
\tilde P &= \frac{\gamma^2}{3}\int_1^\infty
    \frac{(\sqrt{\epsilon^2-1})^3}{e^{\tilde \beta\left(\epsilon-\tilde \mu\right)}+1}\,d\epsilon\,.
\label{eq:pe}
\end{align}
Here we have re-written the integrals in terms of the variable $\epsilon = \sqrt{1+p^2/m^2}$, and we defined the dimensionless coupling $\gamma^2=gGL^2m^4/2\pi^2$.

A convenient re-parametrization of $\lambda$ and $\nu$ in terms of new functions $\tilde{M}$ and $\chi$ is
\begin{eqnarray}
&&e^{\lambda}=\left(1-\frac{2 \tilde{M} }{r}+{r^2}\right)^{-1},	
\label{eq:redefinition-lambda}
\\
&&e^{\nu}=e^\chi\left(1-\frac{2 \tilde{M} }{r}+{r^2}\right)\,.
\label{eq:redefinition-nu}
\end{eqnarray}
The resulting Einstein equations read
\begin{align}
&\frac{d \tilde{M}}{d r}=4\pi\,r^2 \tilde\rho \,,
\label{eq:eqs2a}
\\
%&\frac{d\nu}{d r}=\frac{2}{r^2}\left(M+r^3\left(4\pi\,\tilde P-\frac32%+P_{vac}\right)\right)e^{\lambda(r)}, \\
&\frac{d\chi}{d r}
=8\pi r\left(\tilde P+\tilde\rho\right)e^{\lambda},
%\\
\label{eq:eqs2b}
\end{align}

The equations \eqref{eq:eqs2a}-\eqref{eq:eqs2b}, together with the definitions \eqref{eq:rhoe}-\eqref{eq:pe} in terms of the spatially varying temperature and chemical potentials given by \eqref{eq:tolman}-\eqref{eq:klein} must be solved numerically for the variables $\tilde{M}, \chi,\tilde{\rho},\tilde{P}$. Notice that the fermion mass $m$ plays no role other that setting the scale. We choose boundary conditions at the center of the star $r=0$ as
\begin{eqnarray}
\tilde{M}(0)&=&0\,,\qquad\qquad\mbox{(or in other words }\lambda(0)=0\mbox{)}\,,\nonumber\\
\chi(0)&=&0\,,\nonumber\\
&&\nonumber\\
\tilde T(0)&=&\tilde T_0\,,\nonumber\\
\tilde \mu(0)&=&\tilde \mu_0\ \equiv\ \Theta_0\Tilde T_0+1\,.
\label{mu0}
\end{eqnarray}
Here $\Theta_0$ can be regarded as the central degeneracy. The resulting solutions are indexed by the parameters $\tilde T_0, \Theta_0$ and $\gamma$.

\section{The boundary: degenerate fermionic operators at finite temperature}
\label{sec:correlators}

Since our geometry asymptotes to global AdS, its conformal boundary is a cylinder $\mathbb{R}\times S^2$. The $\mathbb{R}$ direction is coordenatized by the variable $t$ on our metric anzatz. Now, in order to put the system in a thermal bath, we have to go to Euclidean time by the Wick rotation $t_E=it$. Then, we compactify the $t_E$ direction in a circle of length $\beta=e^{\frac{\nu-\nu_0}2}\beta_0$ corresponding to the inverse temperature. Since we want the fluid in the bulk to be in equilibrium with the thermal bath, we need to impose $\beta=1/T$ or, by making use of Tolman relation $\beta_0=1/T_0$. On the aforementioned cylinder, the holographic dual is defined as a conformal field theory. Since the Euclidean time direction of the conformal theory is given by $t_E$, its temperature corresponds to $T_0$.

In order for holography to work, the conformal field theory must have a large central charge. This can be achieved in the present setup without spoiling the perfect fluid approximation, as explained in \cite{deBoer:2009wk}, by taking a proportionally large mass $m$.

In order to probe the above defined background with some conformal field theory observable, we concentrate in the simplest case of a scalar operator.
According to the general dictionary of the AdS/CFT correspondence, the Matsubara two point correlator of a boundary scalar operator with large conformal dimension $\Delta\equiv{\sf m} L$, is given as a function of the angular span of the points on the boundary $\Delta\vartheta, \Delta\varphi$ and the elapsed Euclidean time $\Delta t_E$ as
\begin{equation}
\langle
{\cal O}(\Delta\vartheta, \Delta\varphi, \Delta t_E){\cal O}(0)
\rangle
=
\lim_{r_\epsilon\to\infty}r_\epsilon^{2{\sf m} L}e^{-S^E_{\sf on-shell}(\Delta\vartheta, \Delta\varphi,\Delta t_E)}\,,
\label{eq:correlator}
\end{equation}
where $r_\epsilon$ is an UV bulk regulator, whose power is included in the prefactor in order to get a finite result. The exponent contains the on-shell form of the Euclidean action for a particle with mass ${\sf m}$ moving in the Euclidean bulk
\begin{equation}
S_E ={\sf m} L\int d\tau_E\sqrt{e^{\nu(r)}\, {t'_E}^2+e^{\lambda(r)}\, {r'}^2+r^2({\vartheta'}^2+\sin^2\!\vartheta\,{\varphi'}^2)}\, ,
\label{eq:action-particle}
\end{equation}
where $\tau_E$ is the an Euclidean affine parameter and $(~')=\partial_{\tau_E}(~)$. The worldline approximation was used in \cite{Balasubramanian:2011ur} to study thermalization. For details on its implementation see \cite{Giordano:2014kya} and references therein.

Without loss of generality, we can concentrate in trajectories completely contained in the equatorial plane $\vartheta=\pi/2$.
Moreover, this action is invariant under arbitrary reparametrizations of $\tau_E$, which allows us to fix $\tau_E = \varphi$.
Restricting to trajectories with constant $t_E$ we end up with
\begin{equation}
S_E ={\sf m} L\int d\varphi\sqrt{r^2+e^{\lambda(r)}\,r'^2}\, ,
\label{eq:action-particle-fixed}
\end{equation}
The resulting equations for the single dynamical variable $r$ are invariant under $\varphi$ translations, implying the conservation of the quantity
\begin{equation}
r_*=\frac{\,r^2}{\sqrt{r^2+e^{\lambda(r)}\,r'^2}}\,.
\label{eq:p-phi}
\end{equation}
By evaluating the right hand side at the tip of the trajectory where $r'=0$, we see that $r_*$ corresponds to the radial position of the tip ($r_*(r'=0)=r$) . Solving for $r'$ we get
\begin{equation}
r'=\frac r{r_*}e^{-\frac{\lambda(r)}2}\sqrt{r^2-r_*^2}\,.
\label{eq:velocity}
\end{equation}
This allows us to relate the integration constant $r_*$ with the angular separation at the boundary $\Delta\varphi$ of the starting and final points of the trajectory
\begin{eqnarray}
\Delta \varphi &=&  \int \frac{dr}{r'}
=2r_*\int_{r_*}^{r_\epsilon}\!\!dr \frac{e^{\frac{\lambda(r)}2}}{r\sqrt{r^2-r_*^2}}\,,
\label{eq:anmgular-span}
\end{eqnarray}
where we included a cutoff $r_\epsilon$ in the second line, that must be taken to infinity at the end of the calculations.

On the other hand, plugging eq. \eqref{eq:velocity} into the gauge-fixed action  \eqref{eq:action-particle-fixed} results in the on shell form
\begin{equation}
S^E_{\sf on-shell}(\Delta\varphi) %=\frac{{\sf m}  L}{r_*}\int\!\!d\varphi r^2
=2{\sf m}  L\int_{r_*}^{r_\epsilon}\!\!dr  \frac{re^{\frac{\lambda(r)}2}}{\sqrt{r^2-r_*^2}}
\label{eq:action-particle-on-shell}
\end{equation}
where the same cutoff was included. Notice that, when re-inserted into the correlator \eqref{eq:correlator}, the logarithmic divergence of this integral on the cutoff is canceled by the prefactor, as anticipated.

\section{Results}
\label{sec:results}

\begin{figure}[ht]
\centering
\begin{subfigure}[a]{\textwidth}
\includegraphics[width=.48\textwidth]{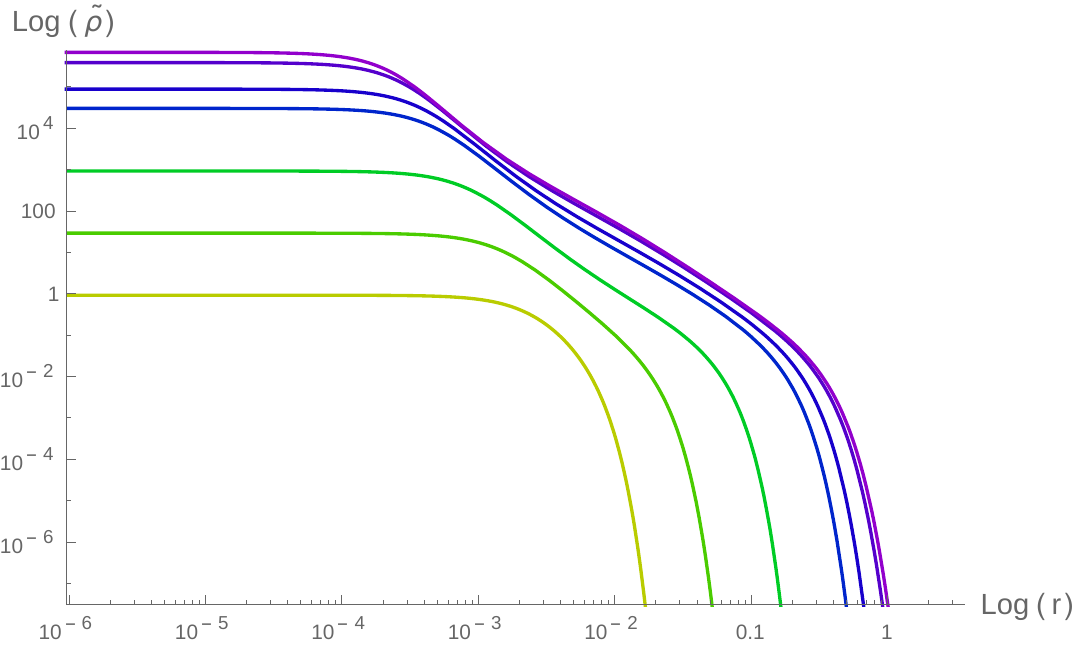}
\hfill
\includegraphics[width=.48\textwidth]{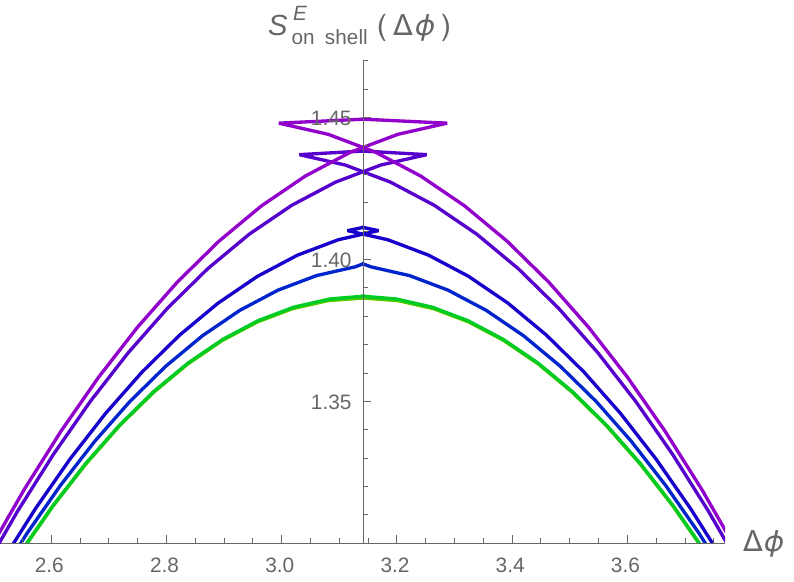}
\end{subfigure}
\\~\\~\\
\begin{subfigure}[b]{\textwidth}
\includegraphics[width=.48\textwidth]{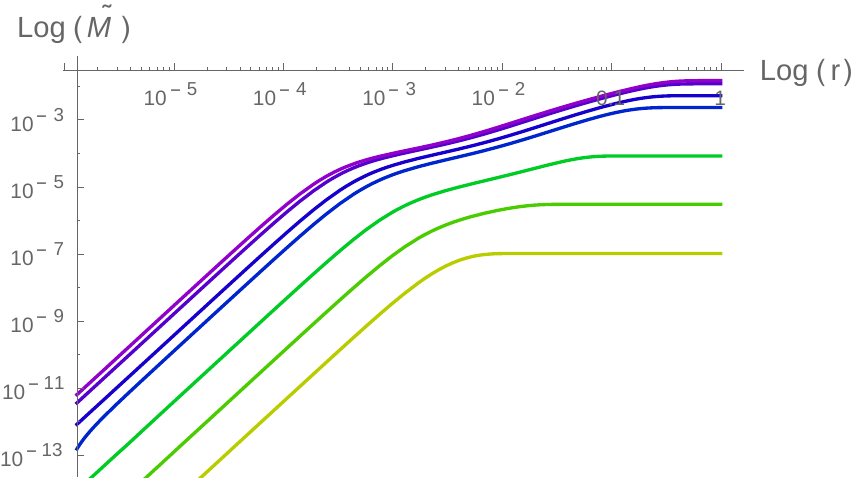}
\quad
\includegraphics[width=.26\textwidth]{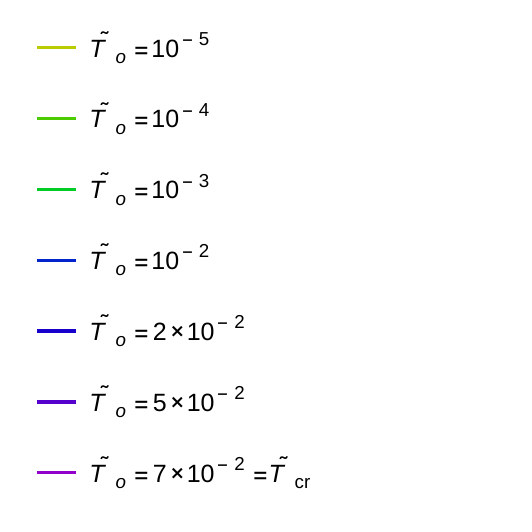}
\end{subfigure}
\put(-60,105){\scriptsize$\Theta_0=-10$\normalsize}
\put(-60,90){\scriptsize$\gamma=7\times10^5$\normalsize}
\caption{
Left: logarithmic plots of the density (up) and mass (down) profiles as functions of the radius, as we approach the critical value of the central temperature (blue curve). Notice that the core and the halo are not well defined for this values of the parameters.
Right (up): plot of the on-shell actions (which correspond to minus the logarithm of the two-point correlator). Notice the swallow tale structure that appears as we approach the critical value for the central temperature { (to make the structure more evident, we extended the natural rank $(0,\pi)$ of the polar angle $\Delta \varphi$ along the opposite meridian up to $2\pi$)}.
\label{fig:thetaneg}
}
\end{figure}

\begin{figure}[ht]
\centering
\begin{subfigure}[a]{\textwidth}
\includegraphics[width=.48\textwidth]{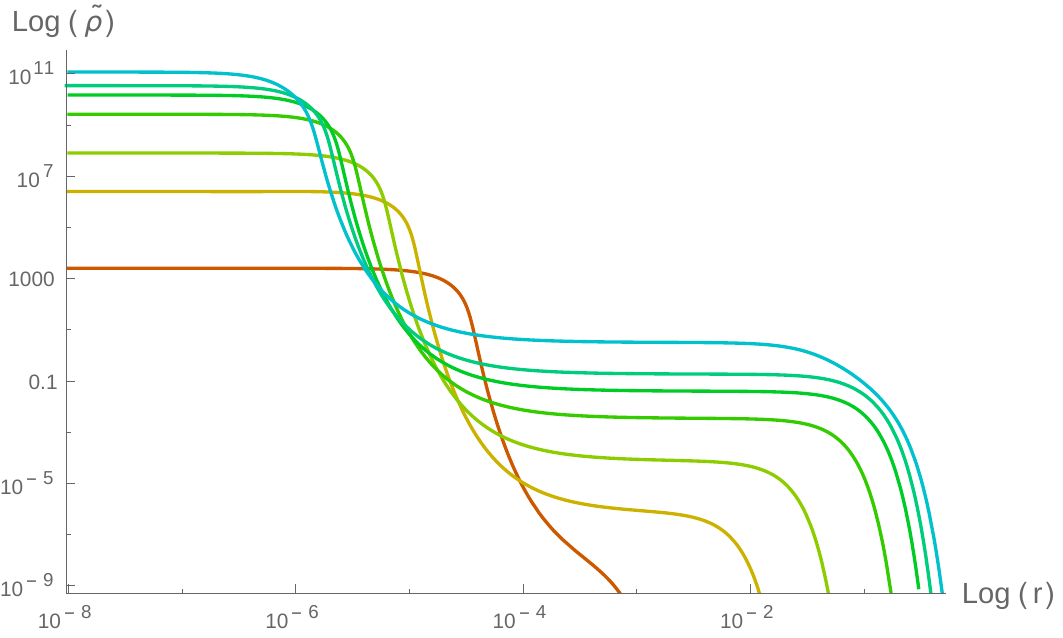}
\hfill
\includegraphics[width=.48\textwidth]{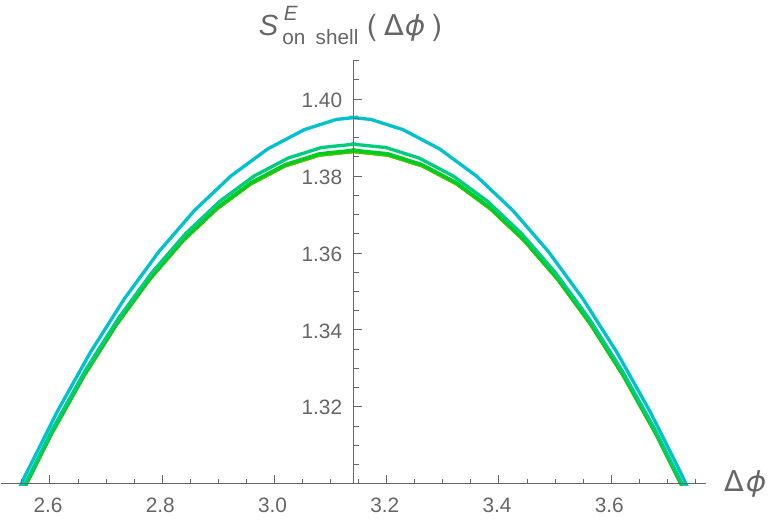}
\end{subfigure}
\\~\\~\\
\begin{subfigure}[b]{\textwidth}
\includegraphics[width=.48\textwidth]{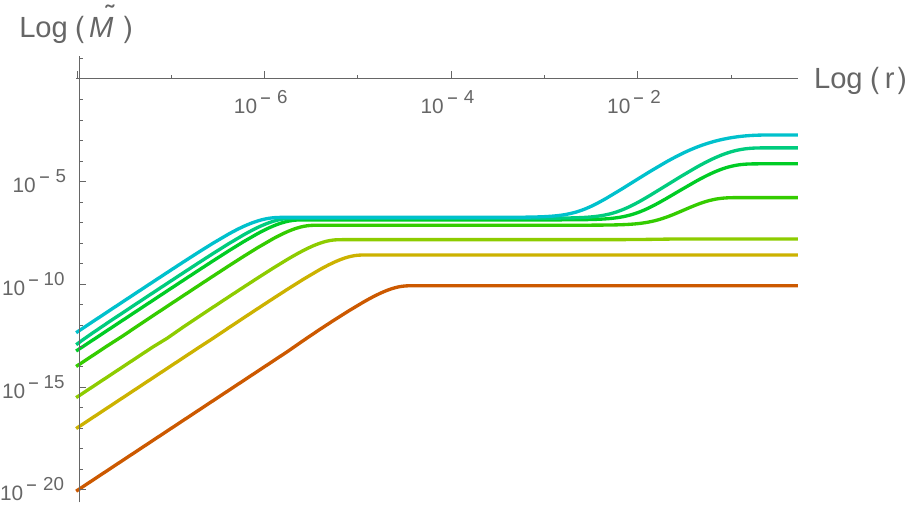}
\quad
\includegraphics[width=.26\textwidth]{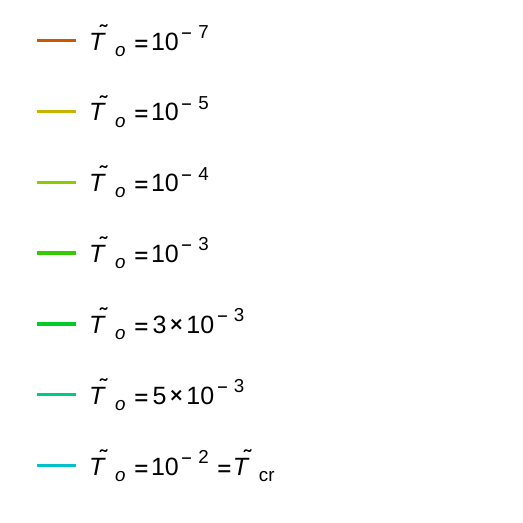}
\end{subfigure}
\put(-60,105){\scriptsize$\Theta_0=30$\normalsize}
\put(-60,90){\scriptsize$\gamma=7\times10^5$\normalsize}
\caption{
Left: logarithmic plots of the density (up) and mass (down) profiles as functions of the radius, as we approach the critical value of the central temperature (blue curve). Notice the that the core and the halo are well defined for this values of the parameters.
Right (up): plot of the on-shell actions (which correspond to minus the logarithm of the two-point correlator) as functions of the angle at the boundary. Notice that there is not a swallow tale structure as we approach the critical value for the central temperature { (to compare with the previous figures, we extended the natural rank $(0,\pi)$ of the polar angle $\Delta \varphi$ along the opposite meridian up to $2\pi$)}.
\label{fig:thetapos}
}
\end{figure}

\begin{figure}[ht]
\centering
\begin{subfigure}[a]{\textwidth}
\includegraphics[width=.48\textwidth]{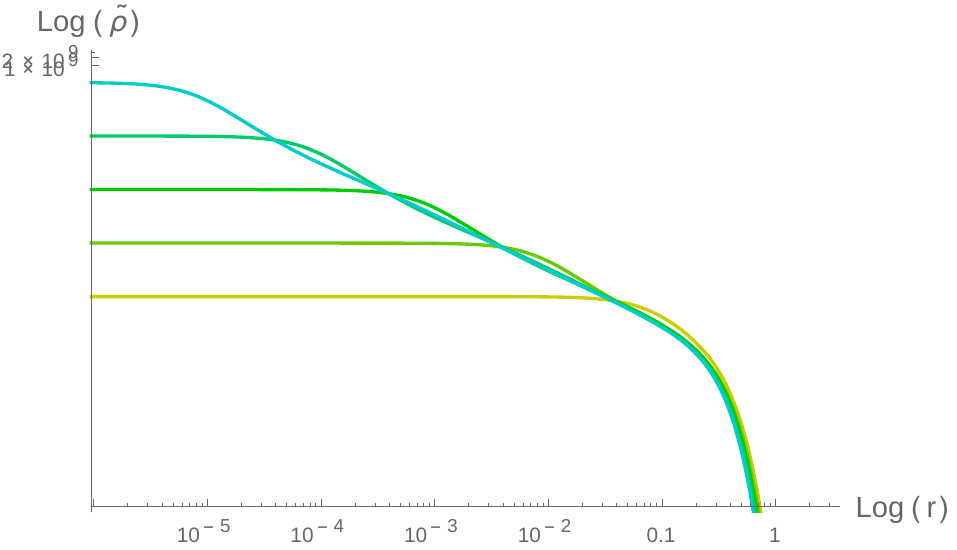}
\hfill
\includegraphics[width=.48\textwidth]{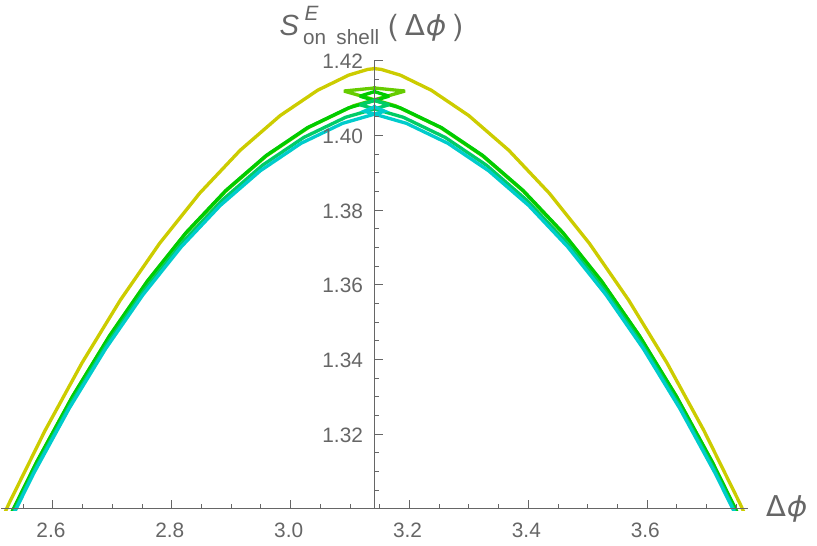}
\end{subfigure}
\\~\\~\\
\begin{subfigure}[b]{\textwidth}
\includegraphics[width=.48\textwidth]{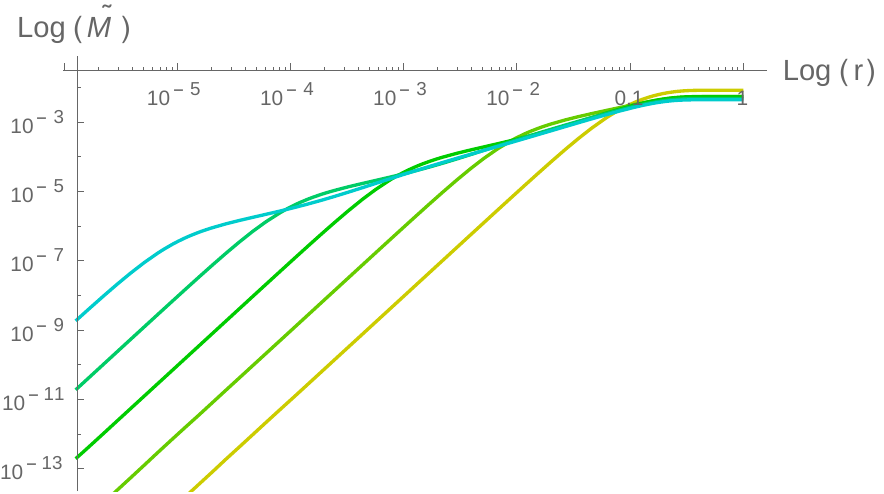}
\quad
\includegraphics[width=.26\textwidth]{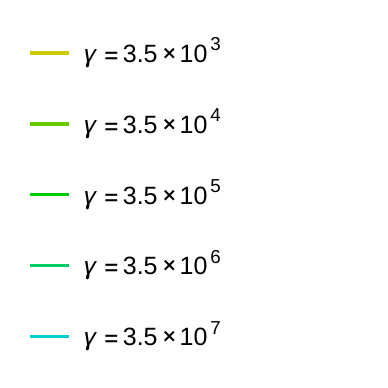}
\end{subfigure}
\put(-60,98){\scriptsize$\Theta_0=-10$\normalsize}
\put(-60,76){\scriptsize$\tilde T_0=2\times10^{-2}$\normalsize}
\caption{
Left: logarithmic plots of the density (up) and mass (down) profiles as functions of the radius, as we move the value of the dimensionless AdS radius $\gamma$. Notice the that the core and the halo are not well defined for this values of the parameters.
Right (up): plot of the on-shell actions (which correspond to minus the logarithm of the two-point correlator) as functions of the angle at the boundary. Notice that a swallow tale structure shows up, as we decrease the value of the dimensionless AdS radius $\gamma$ { (to make the structure more evident, we extended the natural rank $(0,\pi)$ of the polar angle $\Delta \varphi$ along the opposite meridian up to $2\pi$)}.
\label{fig:thetaneggamma}
}
\end{figure}

\begin{figure}[ht]
\centering
\begin{subfigure}[a]{\textwidth}
\includegraphics[width=.48\textwidth]{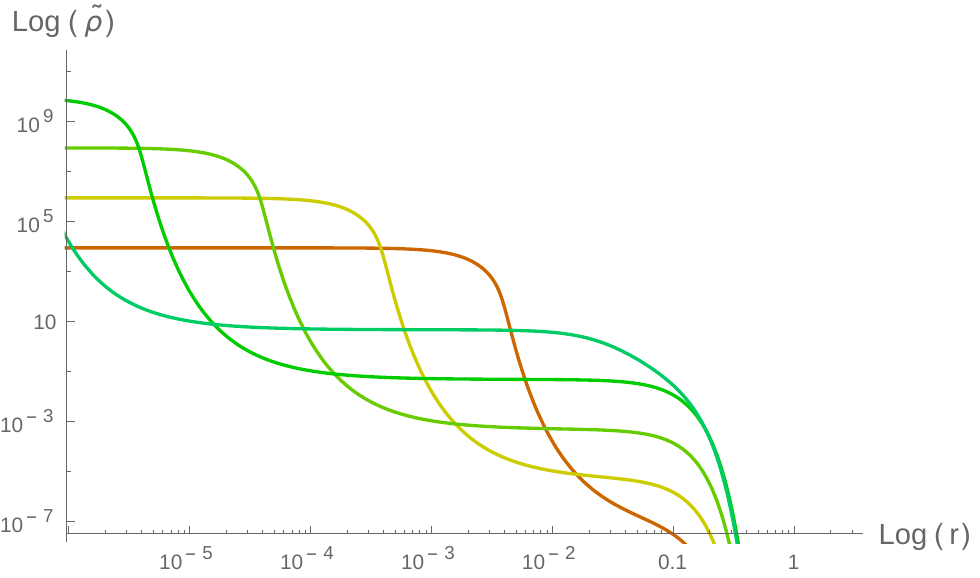}
\hfill
\includegraphics[width=.48\textwidth]{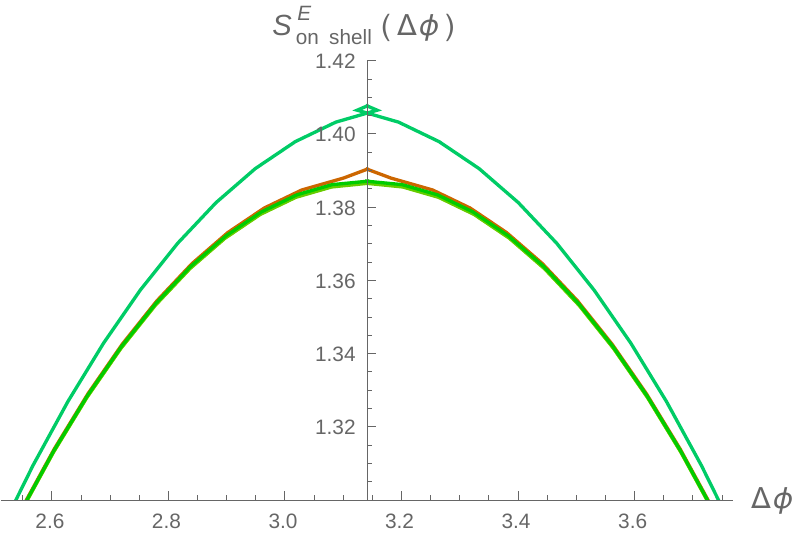}
\end{subfigure}
\\~\\~\\
\begin{subfigure}[b]{\textwidth}
\includegraphics[width=.48\textwidth]{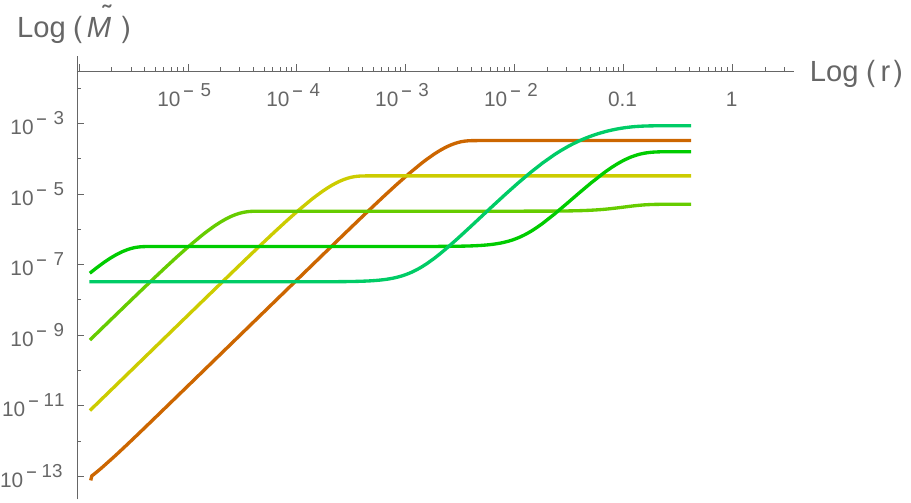}
\quad
\includegraphics[width=.26\textwidth]{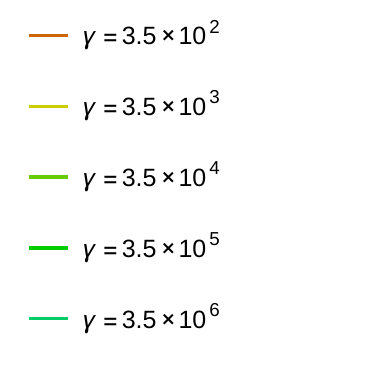}
\end{subfigure}
\put(-60,105){\small$\Theta_0=30$\normalsize}
\put(-60,80){\small$\tilde T_0=5\times10^{-3}$\normalsize}
\caption{
Left: logarithmic plots of the density (up) and mass (down) profiles as functions of the radius, as we move the value of the dimensionless AdS rad\-ius $\gamma$. Not\-ice the that both core and halo are well def\-ined for this values of the parameters.
Right (up): plot of the on-shell actions (which correspond to minus the logarithm of the two-point correlator) as functions of the angle at the boundary. Not\-ice that a small swal\-low tale struc\-ture shows up, as we increase the value of the dimensionless AdS radius $\gamma$ { (to make the structure more evident, we extended the natural rank $(0,\pi)$ of the polar angle $\Delta \varphi$ along the opposite meridian up to $2\pi$)}.~\\~\\~
\label{fig:thetaposgamma}
}
\end{figure}

We solved the system (\ref{eq:redefinition-lambda})-(\ref{mu0}) numerically by using {\tt Fortran} and {\tt Mathematica} routines, for different values of the parameters $ \tilde T_0, \Theta_0$ and $\gamma$. Plots of the resulting density $\tilde{\rho}(r)$ and mass $\tilde{M}(r)$ profiles are shown in Figs. \ref{fig:thetaneg} to \ref{fig:thetaposgamma}.

As can be seen in the plots, the expected core-halo structure appears at large values of $\gamma$ and becomes more marked when the central degeneracy $\Theta_0$ and the central temperature $\tilde T_0$ are large. In those cases, the density profile $\tilde{\rho}(r)$ presents a central plateau that extends up to a well defined radius $r_c$, identified with the ``core'' of the configuration (a precise definition of $r_c$ can be given as the position of the first maximum on the velocity curve), and a lower exterior plateau identified with the ``halo''.
Further in $r$, we see that the density drops to zero at a finite radius $r=r_b$, that defines the radius of the star.

\begin{figure}[ht]
\includegraphics[width=.48\textwidth]{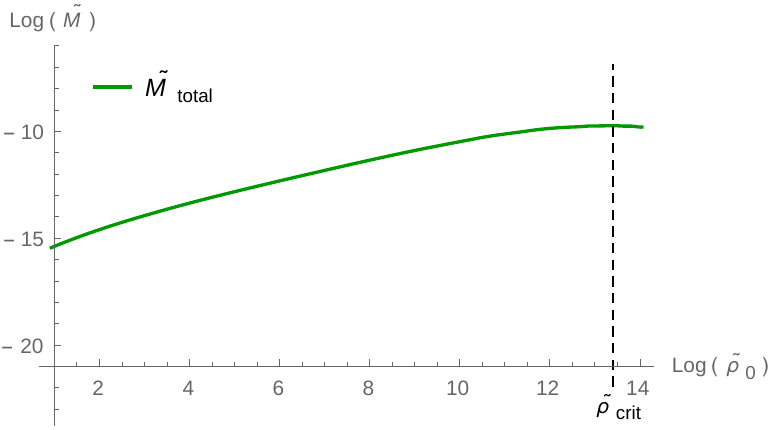}
\hfill
\includegraphics[width=.48\textwidth]{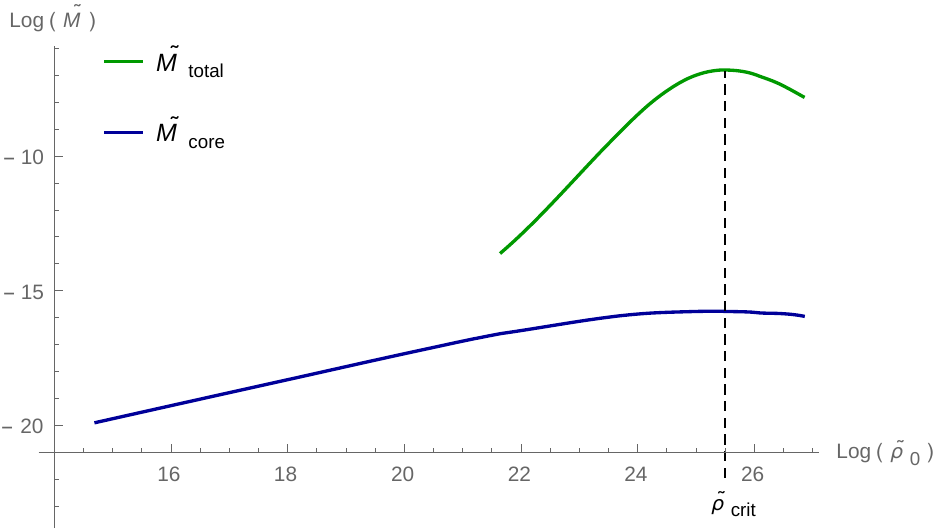}
\put(-365,-10){\scriptsize$\Theta_0=-10$\normalsize}
\put(-130,-10){\scriptsize$\Theta_0=30$\normalsize}
\vspace{.4cm}
\caption{
Logarithmic plots of the mass as a function of the central density, for negative (left) and positive (right) values of the central degeneracy $\Theta_0$. There is a critical value of the central density at which the resulting mass reaches a maximum. For larger values of the central density the star can be regarded as unstable. Notice that, in the case of positive $\Theta_0$ in which the core-halo structure is well differentiated, the core mass has a maximum at the same value of the central density that gives the maximum of the total mass.
\label{fig:critical}}
\end{figure}

The region $r>r_b$ is described by a vacuum Schwarzchild solution with mass $\tilde{M}_b=\tilde{M}(r_b)$, corresponding to the total mass of the star. A plot of the total mass $\tilde{M}_b$ as a function of the central density is shown Fig. \ref{fig:critical}. Similarly to what happens in the asymptotically flat case \cite{Arguelles:2014sfa}, there is a maximum of the mass at a finite critical value of the central density $\tilde{\rho}_{\sf cr}$. Since solutions with higher mass do not exist, this bound is interpreted as the Oppenheimer-Volkoff limit. In the asymptotically flat context, solutions with central densities larger than $\tilde{\rho}_{\sf cr}$ can be shown to be unstable (see \cite{Schiffrin:2013zta} for a recent work). We assume that the same is true in the present asymptotically AdS setup.
Interestingly if we define the core mass as $\tilde{M}_c=\tilde{M}(r_c)$, it has also a maximum for the same value of the central density (see Fig.~\ref{fig:critical}). From the holographic perspective, the gravitational collapse in the bulk can be interpreted as a confinement/deconfinement transition in the dual field theory (analogously to the zero temperature case reported in \cite{Arsiwalla:2011}).

%Fig1

%Fig2

%Fig3

%Fig4

%Fig5
%

%Fig6
\begin{figure}[ht]
\centering
\begin{subfigure}[b]{\textwidth}
\includegraphics[width=.46\textwidth]{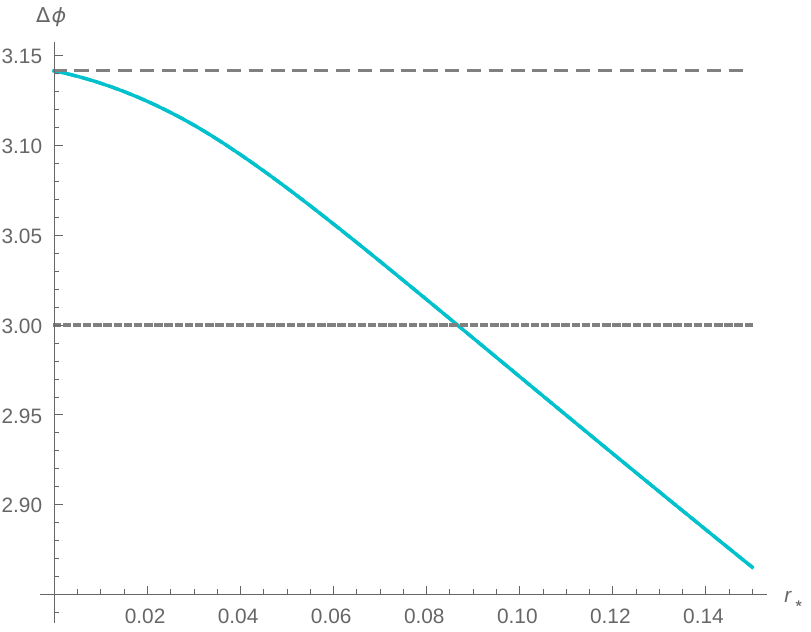}
\hfill
\includegraphics[width=.46\textwidth]{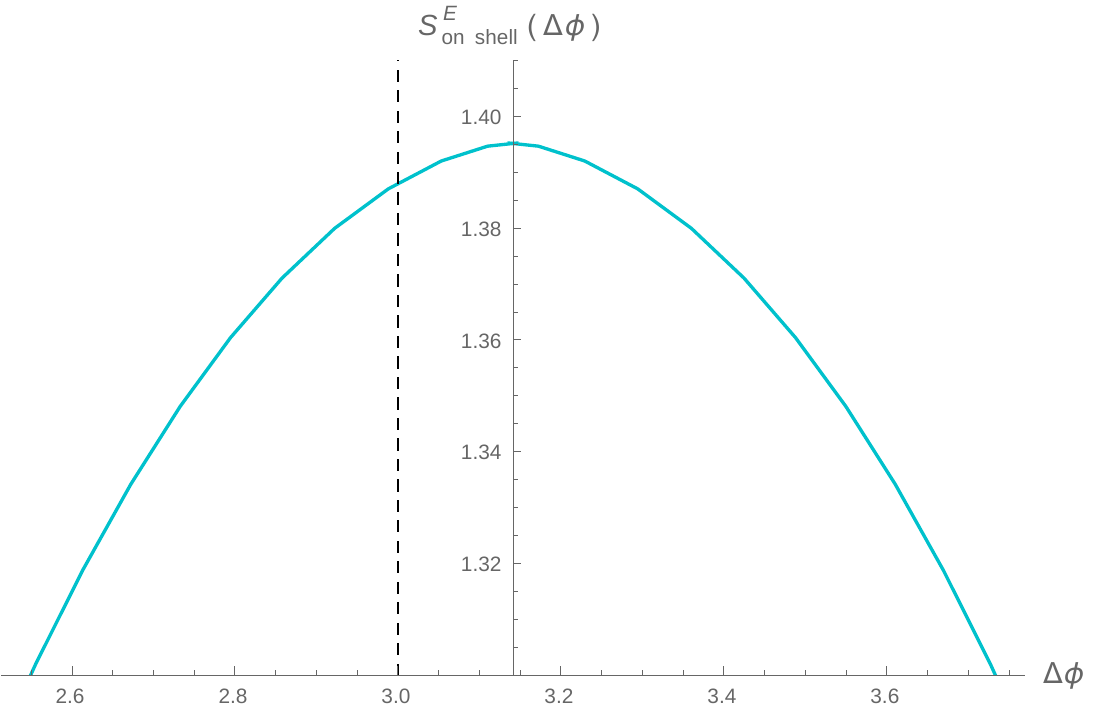}
\end{subfigure} 
\\~\\~\\
\begin{subfigure}[a]{\textwidth}
\includegraphics[width=.46\textwidth]{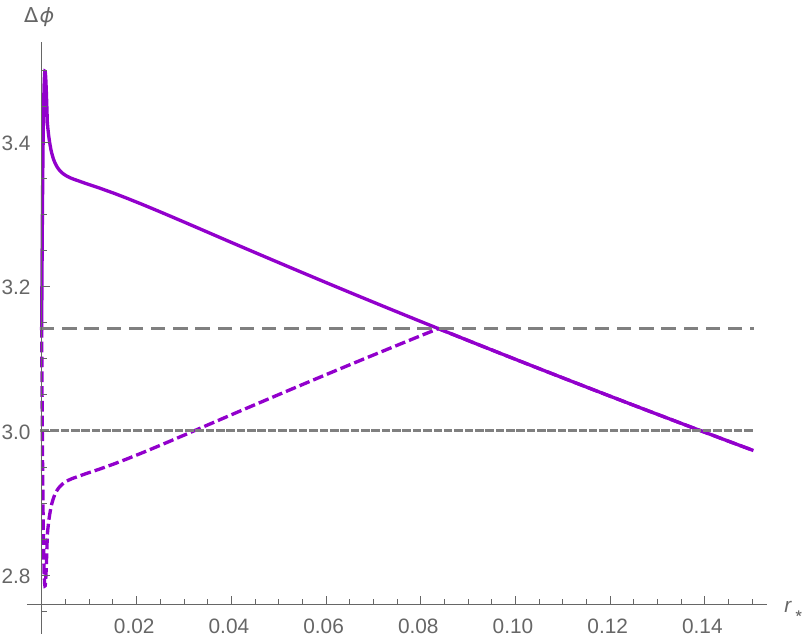}
\hfill
\includegraphics[width=.46\textwidth]{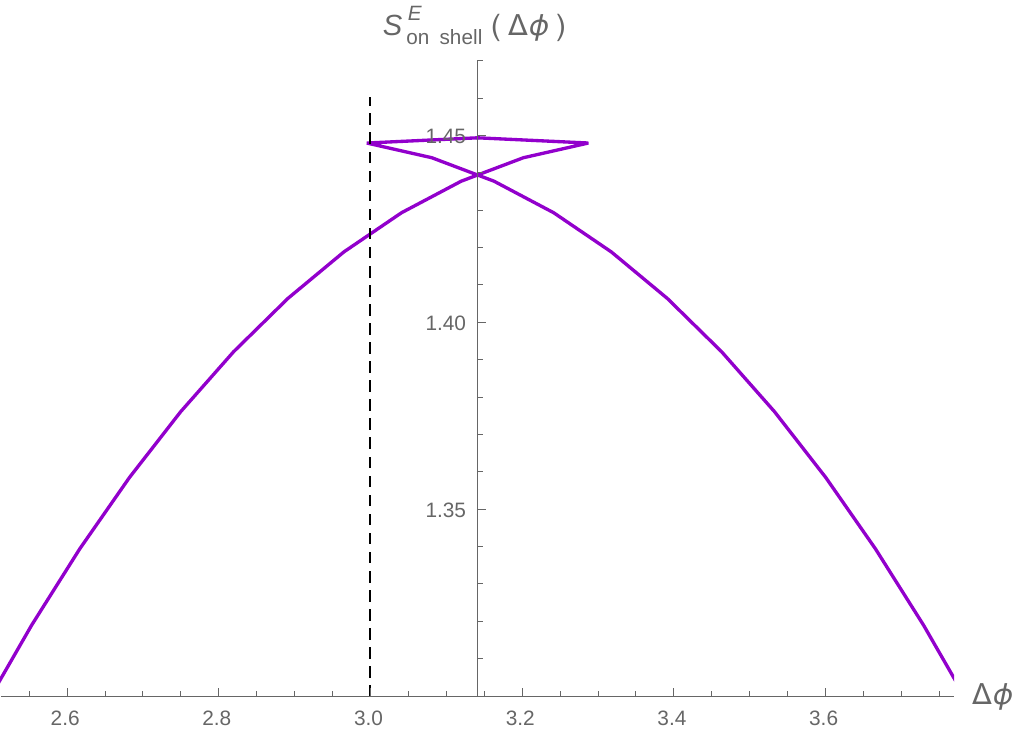}
\end{subfigure}
\caption{
Plots of the angular span $\Delta\varphi$ as a function of the apsidal position $r_*$ (left), and the corresponding correlators (right). Top: plots corresponding to the critical temperature  $T=10^{-2}$ at positive central degeneracy $\Theta_0=30$. The angle $\Delta\varphi$ is a monotonic function of $r_*$, resulting in the univalued correlator (the dotted line sits at $\Delta\varphi=3$ intersects the curves only once).
Bottom: plots corresponding to the critical temperature  $\tilde T=7\times 10^{-2}$ at negative central degeneracy $\Theta_0=-10$. Here the angle oscillates as a function of $r_*$. Since the natural range of the polar angle is $(0,\pi)$, values larger than $\pi$ must be reflected in at $\Delta\varphi=\pi$ (dashed line), resulting in the dashed colored curve. As a result, a given value of the angle corresponds to three values of $r_*$, resulting in a multivalued correlator (the dotted line sits at $\Delta\varphi=3$ intersects the curves three times).
\label{fig:angular}}
\end{figure}
%

%Fig7
\begin{figure}[ht]
\centering
\qquad\qquad
\includegraphics[width=.52\textwidth]{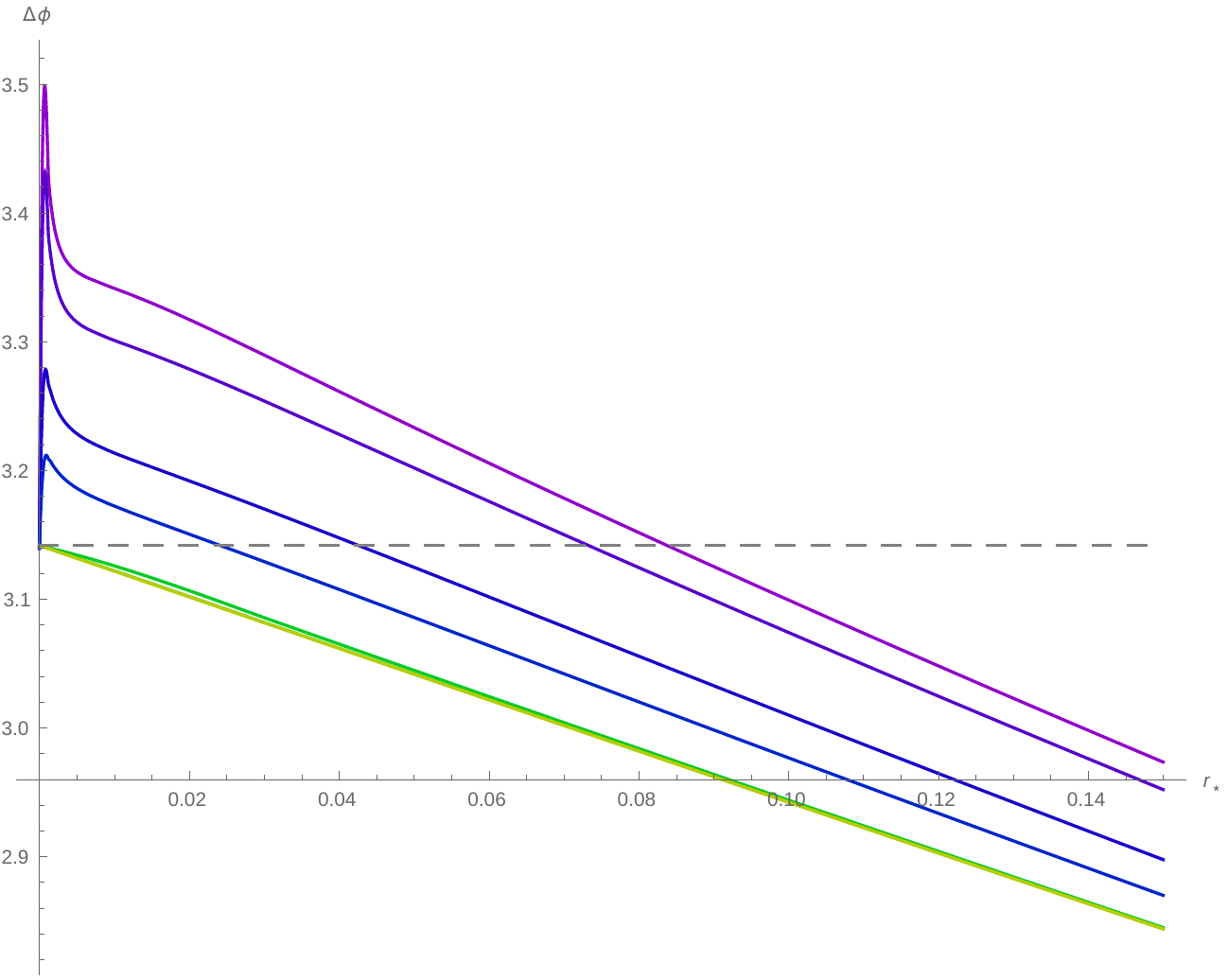}
\hfill
\quad
\includegraphics[width=.35\textwidth]{LegendsNeg.pdf}
\caption{
Plots of the angular span $\Delta\varphi$ as a function of the apsidal position $r_*$. The plots correspond to the temperatures  depicted in Fig. \ref{fig:thetaneg}, at negative central degeneracy $\Theta_0=-10$. For temperatures $\tilde T \gtrsim 10^{-2}$ the angle $\Delta\varphi$ oscillates as a function of $r_*$, resulting in the multivalued correlator. On the other hand, for  $\tilde T \lesssim 10^{-3}$ the angle $\Delta\varphi$ is monotonic and the correlator is univalued. 
\label{fig:hump}}
\end{figure}

Regarding the two point correlators, for clarity in the figures we plotted the on-shell action (which correspond to minus its logarithm) { and we extended the natural rank $(0,\pi)$ of the polar angle $\Delta \varphi$ along the opposite meridian up to $2\pi$}. Interestingly, for solutions near to the critical one, a ``swallow tale'' structure appears. This is evident for small central degeneracies $\Theta_0$, and disappears for larger values. We verified that it is not a cutoff effect, since it remains unchanged when $r_\epsilon$ is moved to infinity. Moreover, if we take the any of the the smooth curves in which we do not see a swallow tale, and zoom in close to $\Delta\varphi=\pi$, we verify that they still smooth, implying that such smoothness is not a finite resolution effect. Remarkably, this effect is more evident for the more massive configurations, \textit{i.e}. configurations with smaller halos and denser cores (see \emph{e.g.} Figs.~\ref{fig:thetaneg} and \ref{fig:thetapos} for comparison).

\newpage

\section{Discussion}

The most interesting feature found in our solutions in the swallow tale structure present in the correlator. Such ``multivalued'' form for correlators have been reported previously for quenched states in thermalization studies, when the correlator is plotted as a function of the gauge theory time \cite{Balasubramanian:2011ur}-\cite{Giordano:2014kya}. Here, it shows up for near to critical equilibrium states, as a function of the spatial separation. 
 When a swallow tale appears in the plots of the free energy as a function of temperature, it is taken as a signal of a phase transition. In our case, such conclusion cannot be drawn without further analysis, that we perform in this section.

From the bulk perspective, the swallow tale structure corresponds to the existence of three different geodesics joining the same points at the boundary.  One of this solutions represents a maximum of the action, the second one represents a saddle point, while the remaining one corresponding to the true minimum. Since the correct value of the dual field theory correlator is given by the curve corresponding to the minimum action, only the lower curve should be kept, and this swallow tale structure corresponds to a discontinuity in the angular derivative of the correlator at antipodal points $\Delta\varphi=\pi$. 
 
Recall that the angular span $\Delta\varphi$ and the on shell action $S^E_{\sf on~shell}$ are obtained from equations \eqref{eq:anmgular-span} and \eqref{eq:action-particle-on-shell} respectively, as a function of the radial position of the apsidal point of the geodesic $r_*$. Then the curve $S^E_{\sf on~shell}$ {\it vs.} $\Delta\varphi$ is plotted using $r_*$ as an affine parameter. In the cases in which the correlator shows a swallow tale, three values of the on-shell action correspond to the same value of the angle. Since the on-shell action is a well defined function of $r_*$, we conclude that there must be three different values of $r_*$ that return the same value of the angle. In other words, the curve $\Delta\varphi$ versus $r_*$ has an oscillation. 

In order to have an oscillation, the derivative of $\Delta\varphi$ with respect to $r_*$ must change its sign. Rewriting equation \eqref{eq:anmgular-span} in terms of the variable $u=r/r_*$ as
\begin{eqnarray}
\Delta \varphi &=&  
2\int_{1}^{\infty}\!\!du \,\frac{e^{\frac{\lambda(r_*u)}2}}{u\sqrt{u^2-1}}\,,
\label{eq:anmgular-span-u}
\end{eqnarray}
we can take the derivative with respect to $r_*$, obtaining%\footnote{Here we are assuming that we can interchange differentiation and integration. }
\begin{eqnarray}
\frac{\partial\Delta\varphi }{\partial r_*} &=&  2
\int_{1}^{\infty}\!\!du \,\frac{\left(e^{\frac{\lambda(r_*u)}2}\right)'}{\sqrt{u^2-1}}\,=
%\nonumber\\
%&=&
2\int_{r_*}^{\infty}\!\!dr \,\frac{e^{\frac{3\lambda(r)}2}}{\sqrt{r^2-r_*^2}}
\left(
\frac{\tilde M'}r-\frac{\tilde M}{r^2}-r
\right)\,,
\label{eq:derivative-anmgular-span-u}
\end{eqnarray}
where in the last equality we used \eqref{eq:redefinition-lambda} and reintroduced the variable $r$. This can be rearranged as
\begin{eqnarray}
\frac{\partial\Delta\varphi }{\partial r_*} &=&  
2\int_{r_*}^{\infty}\!\!dr \,\frac{e^{\frac{3\lambda(r)}2}}{\sqrt{r^2-r_*^2}}
\,p\,,
\quad\quad\ \quad
{\rm with}
\ \
p=
\left(
\frac{\tilde{M}}{r^2}
\left(\frac{d\log \tilde{M}}{d\log r}-1\right)-r
\right)\,,
\label{eq:derivative-anmgular-span-u-log}
\end{eqnarray}
In order to understand whether the $\Delta \varphi$ oscillates as a function of $r_*$, we need to study whether the above integral changes its sign. The sign of the integrand is fixed by the sign of $p$. The overall sign of the integral for a given value of $r_*$ is then determined by the relative weight of the regions in which $p$ is positive or negative. 

For a radius $r_*$ larger than the radius of the star $r_*>r_b$, the mass function is constant $\tilde{M}=\tilde{M}_b$. Then its logarithmic derivative in $p$ vanishes, resulting in a negative overall value for the integral. This implies ${\partial\Delta\varphi }/{\partial r_*}<0$,  in agreement with the intuition that the further in the interior of the geometry our geodesic reaches, the larger the angular span of the initial and final points in the boundary.

As we move $r_*$ into the halo $r_c<r_*<r_b$, such region starts to contribute to the integral. Since at the halo the density is approximately constant $\tilde{\rho}\simeq\tilde{\rho}_h$ and very small, the mass function takes the  form $\tilde{M}=\tilde{M}_c+{4\pi}\tilde{\rho}_h r^3/3$, that results in $p\simeq 2\tilde{M}_c/r^2-r$.
For light configurations, the first term is small and then $p$ is negative, giving a negative contribution to the integral (this is the case for example for the critical case in Fig. \ref{fig:thetapos}). On the other hand, for dense cores, it results in a positive contribution (as is the case for the critical curve in Fig. \ref{fig:thetaneg}).
Whether such contribution can change the overall sign of the integral to ${\partial\Delta\varphi }/{\partial r_*}>0$ depends on the density of the configuration.

When $r_*$ gets into the core $r_*<r_c$, the mass function can be written as $\tilde{M}=4\pi \tilde{\rho}_c r^3/3$, where $\rho_c$ is the approximately constant core density. We get $p=(8\pi\tilde{\rho}_c-3)r/3$, which is positive for the central densities we considered, resulting in a positive contribution to the integral. Again, such positive contribution may or may not be enough to change the sign of the overall integral, depending of the density of the background.

The conclusion is that a negative derivative ${\partial\Delta\varphi }/{\partial r_*}<0$ is always the case for large $r_*$. On the other hand, for small $r_*$ the derivative still negative for light configurations, but it may become positive if the configuration is massive enough. In order for this to happen, we need that the positive contribution extends for a large enough interval of the radial variable, {\em i.e.} we need an extended core.  
If such is the case, we have an oscilation of $\Delta\varphi$ as a function of $r_*$, that results in a swallow tale in the correlator. See Fig. \ref{fig:angular} for further details.

As this discussion should make clear, the swallow tale is not a feature of all correlators that grows continuously as we move the parameters, but a distinctive characteristic of dense and extended cores, which is absent in light ones. In other words, the existence or not of a swallow tale sets a frontier on the $(\Theta_0,\tilde T)$ plane, and in such (somewhat limited) sense, it can be regarded as a signal of a phase transition. For example, at negative $\Theta_0=-10$, we see in Fig. \ref{fig:thetaneg} that the transition temperature at which the swallow tale appears is around $T^{-3}\lesssim\tilde T\lesssim 10^{-2}$. In Fig. \ref{fig:hump} we plotted the corresponding $\Delta\varphi$ curves, showing how the oscillation dissapears as the temperature gets smaller.

\newpage
\section{Conclusions}
\label{sec:concusions}
We investigated the solution to Einstein equations sourced by to a weakly-coupled self-gravitating fermionic fluid at finite temperature in global AdS spacetime. They form a three-parameter family, indexed by the central degeneracy, the central temperature and the particle mass. According to the value of such parameters, a well defined core-halo structure appears when the mass is large, and it becomes better defined as the central degeneracy grows and the central temperature decreases.

With the obtained backgrounds, we evaluated the two point correlator of a scalar operator of the dual field theory at constant time. We found that it shows a swallow tale structure when evaluated on near to critical solutions with small central degeneracy. This is to be compared to a similar phenomenon previously reported in the case of non-equilibrium quenched states. The presence or absence of such swallow tale sets a frontier in the phase diagram, that allows us to identify separated regions. In such limited sense, the appearance of a swallow tale can be interpreted as a phase transition. 

A possible future line of research is the comparison of free energies between the present configuration for finite temperature matter and the black hole solution. Another possibility is the extension of the present results to a charged fluid in the Poincare patch of AdS, the so-called electron star solution.

\section{Acknowledgments}

This works was partially supported by grants PIP-2008-0396 (Conicet, Argentina) and PID-2013-X648 and PID-2017-X791 (UNLP, Argentina).  We thank Guillermo Silva, Mauricio Sturla, Pablo Pisani, Gast\'on Giribet, Daniela D'Ascanio and Mariel Santangelo for helpful discussions regarding the interpretation of the swallow tale structure on the boundary correlation function.


\begin{thebibliography}{99}

%\cite{Faulkner:2009wj}
\bibitem{Faulkner:2009wj}
  T.~Faulkner, H.~Liu, J.~McGreevy and D.~Vegh,
  ``Emergent quantum criticality, Fermi surfaces, and AdS(2),''
  Phys.\ Rev.\ D {\bf 83} (2011) 125002
  doi:10.1103/PhysRevD.83.125002
  [arXiv:0907.2694 [hep-th]].
  %%CITATION = doi:10.1103/PhysRevD.83.125002;%%
  %413 citations counted in INSPIRE as of 12 Jul 2017

%\cite{Lee:2008xf}
\bibitem{Lee:2008xf}
  S.~S.~Lee,
  ``A Non-Fermi Liquid from a Charged Black Hole: A Critical Fermi Ball,''
  Phys.\ Rev.\ D {\bf 79} (2009) 086006
  doi:10.1103/PhysRevD.79.086006
  [arXiv:0809.3402 [hep-th]].
  %%CITATION = doi:10.1103/PhysRevD.79.086006;%%
  %268 citations counted in INSPIRE as of 12 Jul 2017

%\cite{deBoer:2009wk}
\bibitem{deBoer:2009wk}
  J.~de Boer, K.~Papadodimas and E.~Verlinde,
  ``Holographic Neutron Stars,''
  JHEP {\bf 1010} (2010) 020
  doi:10.1007/JHEP10(2010)020
  [arXiv:0907.2695 [hep-th]].
  %%CITATION = doi:10.1007/JHEP10(2010)020;%%
  %46 citations counted in INSPIRE as of 12 Jul 2017

\bibitem{Arsiwalla:2011}
  X.~Arsiwalla, J.~de Boer, K.~Papadodimas and E.~Verlinde,
  ``Degenerate stars and gravitational collapse in AdS/CFT,''
   JHEP {\bf 01} (2011) 144
  doi:10.1007/JHEP01(2011)144
  [arXiv:1010.5784 [hep-th]].


%\cite{Hartnoll:2010gu}
\bibitem{Hartnoll:2010gu}
  S.~A.~Hartnoll and A.~Tavanfar,
  ``Electron stars for holographic metallic criticality,''
  Phys.\ Rev.\ D {\bf 83} (2011) 046003
  doi:10.1103/PhysRevD.83.046003
  [arXiv:1008.2828 [hep-th]].
  %%CITATION = doi:10.1103/PhysRevD.83.046003;%%
  %132 citations counted in INSPIRE as of 12 Jul 2017

%\cite{Hartnoll:2011dm}
\bibitem{Hartnoll:2011dm}
  S.~A.~Hartnoll, D.~M.~Hofman and D.~Vegh,
  ``Stellar spectroscopy: Fermions and holographic Lifshitz criticality,''
  JHEP {\bf 1108} (2011) 096
  doi:10.1007/JHEP08(2011)096
  [arXiv:1105.3197 [hep-th]].
  %%CITATION = doi:10.1007/JHEP08(2011)096;%%
  %74 citations counted in INSPIRE as of 12 Jul 2017

%\cite{Cubrovic:2009ye}
\bibitem{Cubrovic:2009ye}
  M.~Cubrovic, J.~Zaanen and K.~Schalm,
  %``String Theory, Quantum Phase Transitions and the Emergent Fermi-Liquid,''
  Science {\bf 325} (2009) 439
  doi:10.1126/science.1174962
  [arXiv:0904.1993 [hep-th]].
  %%CITATION = doi:10.1126/science.1174962;%%
  %330 citations counted in INSPIRE as of 15 Dec 2017

%\cite{Hartnoll:2010ik}
\bibitem{Hartnoll:2010ik}
  S.~A.~Hartnoll and P.~Petrov,
  %``Electron star birth: A continuous phase transition at nonzero density,''
  Phys.\ Rev.\ Lett.\  {\bf 106} (2011) 121601
  doi:10.1103/PhysRevLett.106.121601
  [arXiv:1011.6469 [hep-th]].
  %%CITATION = doi:10.1103/PhysRevLett.106.121601;%%
  %43 citations counted in INSPIRE as of 12 Mar 2018
  
%\cite{Puletti:2010de}
\bibitem{Puletti:2010de}
  V.~G.~M.~Puletti, S.~Nowling, L.~Thorlacius and T.~Zingg,
  ``Holographic metals at finite temperature,''
  JHEP {\bf 1101} (2011) 117
  doi:10.1007/JHEP01(2011)117
  [arXiv:1011.6261 [hep-th]].
  %%CITATION = doi:10.1007/JHEP01(2011)117;%%
  %39 citations counted in INSPIRE as of 12 Jul 2017

\bibitem{Gao}  Gao, J. G., Merafina, M.,  Ruffini, R. (1990). ``The semidegenerate configurations of a selfgravitating system of fermions''. Astronomy and Astrophysics, 235, 1-7.

\bibitem{Bilic2002}
N.~Bilic, F.~Munyaneza, G.~B.~Tupper, and R.~D~Viollier,
``The dynamics of stars near Sgr A* and dark matter at the center and in the halo of the galaxy''
Progress in Particle and Nuclear Physics, 48, 291-300 (2002)
doi:10.1016/S0146-6410(02)00136-9

\bibitem{Arguelles2013}
C.~R.~Argüelles, I.~Siutsou, R.~Ruffini, J.~A.~ Rueda, B.~Machado,
``On the core-halo constituents of a semi-degenerate gas of massive fermions''
BAAS, Probes of Dark Matter on Galaxy Scales, 45, 30204 (2013).
%[arXiv:1402.0700 [astro-ph.GA]]

\bibitem{ArguellesJKPS2014}
C.~R.~Argüelles, R.~Ruffini, I.~Siutsou, B.~Fraga, B.
``On the distribution of dark matter in galaxies: quantum treatments.''
Journal of the Korean Physical Society, 65(6), 801-804 (2014).
[arXiv:1402.0700 [astro-ph.GA]]

\bibitem{Ruffini2015}
R.~Ruffini, C.~R.~Argüelles, and J.~A.~ Rueda
``On the core-halo distribution of dark matter in galaxies.''
Monthly Notices of the Royal Astronomical Society 451.1 (2015): 622-628.
[arXiv:1409.7365 [astro-ph.GA]]

\bibitem{ArguellesJCAP2016}
C.~R.~Argüelles, N.~E.~Mavromatos,  J.~A.~Rueda,  R.~Ruffini,
``The role of self-interacting right-handed neutrinos in galactic structure.''
JCAP 2016.04 (2016): 038.
[arXiv:1502.00136 [astro-ph.GA]]

\bibitem{ArguellesMNRAS2016}
C.~R.~Argüelles, A.~Krut, J.~A.~Rueda, R.~Ruffini,
``Novel constraints on fermionic dark matter from galactic observables.''
MNRAS (submitted) (2016).
[arXiv:1502.00136 [astro-ph.GA]]

%\cite{Balasubramanian:2011ur}
\bibitem{Balasubramanian:2011ur}
  V.~Balasubramanian {\it et al.},
  %``Holographic Thermalization,''
  Phys.\ Rev.\ D {\bf 84} (2011) 026010
  doi:10.1103/PhysRevD.84.026010
  [arXiv:1103.2683 [hep-th]].
  %%CITATION = doi:10.1103/PhysRevD.84.026010;%%
  %243 citations counted in INSPIRE as of 15 Dec 2017

%\cite{Giordano:2014kya}
\bibitem{Giordano:2014kya}
  A.~Giordano, N.~E.~Grandi and G.~A.~Silva,
  %``Holographic thermalization of charged operators,''
  JHEP {\bf 1505} (2015) 016
  doi: 10.1007/ JHEP05 (2015) 016
  [arXiv:1412.7953 [hep-th]].
  %%CITATION = doi:10.1007/JHEP05(2015)016;%%
  %7 citations counted in INSPIRE as of 18 Oct 2017

%\cite{Arguelles:2014sfa}
\bibitem{Arguelles:2014sfa}
  C.~R.~Argüelles, R.~Ruffini and B.~M.~O.~Fraga,
  %``Critical configurations for a system of semidegenerate fermions,''
  J.\ Korean Phys.\ Soc.\  {\bf 65} (2014) no.6,  809
  doi:10.3938/jkps.65.809
  [arXiv:1402.1329 [astro-ph.GA]].
  %%CITATION = doi:10.3938/jkps.65.809;%%
  %5 citations counted in INSPIRE as of 22 Dec 2017

%\cite{Schiffrin:2013zta}
\bibitem{Schiffrin:2013zta}
  J.~S.~Schiffrin and R.~M.~Wald,
  %``Turning Point Instabilities for Relativistic Stars and Black Holes,''
  Class.\ Quant.\ Grav.\  {\bf 31} (2014) 035024
  doi:10.1088/0264-9381/31/3/035024
  [arXiv:1310.5117 [gr-qc]].
  %%CITATION = doi:10.1088/0264-9381/31/3/035024;%%
  %8 citations counted in INSPIRE as of 15 Dec 2017
  
%\cite{Solodukhin:1998ec}
\bibitem{Solodukhin:1998ec}
  S.~N.~Solodukhin,
  %``Correlation functions of boundary field theory from bulk Green's functions and phases in the boundary theory,''
  Nucl.\ Phys.\ B {\bf 539} (1999) 403
  doi:10.1016/S0550-3213(98)00715-9
  [hep-th/9806004].
  %%CITATION = doi:10.1016/S0550-3213(98)00715-9;%%
  %20 citations counted in INSPIRE as of 13 Mar 2018
  
\end{thebibliography}
\end{document}